\documentclass[aps,reprint,showpacs,showkeys,preprintnumbers,superscriptaddress,amsmath,amssymb,showkeys]{revtex4-1}

\usepackage{graphicx}
\usepackage{dcolumn}
\usepackage{bm}
\usepackage{hyperref}
\usepackage[dvips]{color}

\begin{document}


\preprint{}


\title{Disorder and magnetic field induced Bose-metal state in two-dimensional Ta$_x$(SiO$_2$)$_{1-x}$ granular films}


\author{Zhi-Hao He}
\affiliation{Tianjin Key Laboratory of Low Dimensional Materials Physics and
Preparing Technology, Department of Physics, Tianjin University, Tianjin 300354,
China}
\author{Hua-Yao Tu}
\affiliation{National Laboratory for Infrared Physics, Shanghai Institute of Technical Physics, Chinese Academy of Science, Shanghai 200083,
China}
\author{Kuang-Hong Gao}
\affiliation{Tianjin Key Laboratory of Low Dimensional Materials Physics and
Preparing Technology, Department of Physics, Tianjin University, Tianjin 300354,
China}
\author{Guo-Lin Yu}
\affiliation{National Laboratory for Infrared Physics, Shanghai Institute of Technical Physics, Chinese Academy of Science, Shanghai 200083,
China}
\author{Zhi-Qing Li}
\email[Corresponding author, e-mail: ]{zhiqingli@tju.edu.cn}
\affiliation{Tianjin Key Laboratory of Low Dimensional Materials Physics and
Preparing Technology, Department of Physics, Tianjin University, Tianjin 300354,
China}



\date{\today}

\begin{abstract}
The origin of the intermediate anomalous metallic state in two-dimensional superconductor materials remains enigmatic. In the present paper, we observe such a state in a series of $\sim$9.0\,nm thick Ta$_x$(SiO$_2$)$_{1-x}$ ($x$ being the volume fraction of Ta) nanogranular films. At zero field, the $x$ $\gtrsim$ 0.75 films undergo a Berezinskii-Kosterlitz-Thouless transition as transform from normal to superconducing states upon cooling. For the $x$ $\lesssim$ 0.71 films, the resistance increases with decreasing temperature from 2 K down to 40\,mK. A normal state to anomalous metallic state transition is observed in the $x$ $\simeq$ 0.73 film, i.e., near the transition temperature, the resistance of the film decreases sharply upon cooling as if the system would cross over to superconducting state, but then saturates to a value far less than that in normal state. When a small magnetic field perpendicular to the film plane is applied, the anomalous metallic state occurs in the  $x$ $\gtrsim$ 0.75 films. It is found that both disorder and magnetic field can induce the transition from superconductor to anomalous metal and their influences on the transition are similar. For the the magnetic field induced case, we find the sheet resistance $R_{\square}(T,H)$ ($T$ and $H$ being the temperature and the magnitude of magnetic field) data near the crossover from the anomalous metal to superconductor and in the vicinity of the anomalous metal to insulator transition, respectively, obey unique scaling laws deduced from the Bose-metal model. Our results strongly suggest that the anomalous metallic state in the Ta$_x$(SiO$_2$)$_{1-x}$ granular films is bosonic and dynamical gauge field fluctuation resulting from superconducting quantum fluctuations plays a key role in its formation.
\end{abstract}


\maketitle


\section{Introduction}
Recently, the anomalous metallic state in two-dimensional (2D) superconductors has attracted great attention~\cite{no1,no2,no3,no4,no5,no6,no7,no8,no9,no10,no11,no12,no13,no14,no15,no16,no17,no18,no19,no20}. The main characteristic of the anomalous metal state is that when the sample is cooled down to a certain temperature, the resistance will drop dramatically as if the system were approaching a superconducting ground state, but then saturates to a value that can be orders of magnitude smaller than that of normal state with further decreasing temperature. The anomalous metallic state has been observed in disordered 2D superconductor films~\cite{no1,no2,no3,no4,no5,no6}, mechanical exfoliated crystalline 2D superconductors~\cite{no9,no10,no11}, the electric-double-layer transistor~\cite{no7,no8}, and Josephson junction array~\cite{no14,no15,no16}, and can be tuned by disorder, magnetic field, and gate voltage. Several theoretical frameworks, including Bose-metal model~\cite{no21,no22,no23,no24}, vortex tunneling model~\cite{no25}, binary composite structure model with superconducting grains embedded in a normal-metal film~\cite{no20,no26,no27}, and new two-fluid model~\cite{no28}, have been proposed to disclose the nature of the intermediate metallic state. However, its origin is still intensely debate. In the Bose-metal and vortex tunneling models, the systems  are both assumed to be composed of superconducting grains and insulating matrix. Thus the superconductor-insulator granular composites are important model system for studying the anomalous metallic state. In fact, the disorder driven anomalous metallic state has appeared in the earlier literatures of 2D granular superconductors and the did not received much attention at that time~\cite{no29,no30}. Thus it is necessary to explore the nature of the anomalous metallic state in superconductor-insulator granular system.

In this paper, we systematically investigate the low-temperature electrical transport properties of a series of Ta$_x$(SiO$_2$)$_{1-x}$ (with $x$ being the volume fraction of Ta) granular films with thickness $\sim$9.0\,nm. One reason for choosing Ta as the superconductor constitute is Ta possesses high
resistance to oxidation at room temperature, another is the anomalous metallic state has been observed in ultrathin pure Ta films~\cite{no3,no4}. It is found that the anomalous metallic state appears in the Ta$_x$(SiO$_2$)$_{1-x}$ films and can not only be tuned by disorder but also by magnetic field. The scaling relation in resistance with magnetic field and temperature can be described by the prediction of Bose-metal scenario.

\begin{figure}
\begin{center}
\includegraphics[scale=1.1]{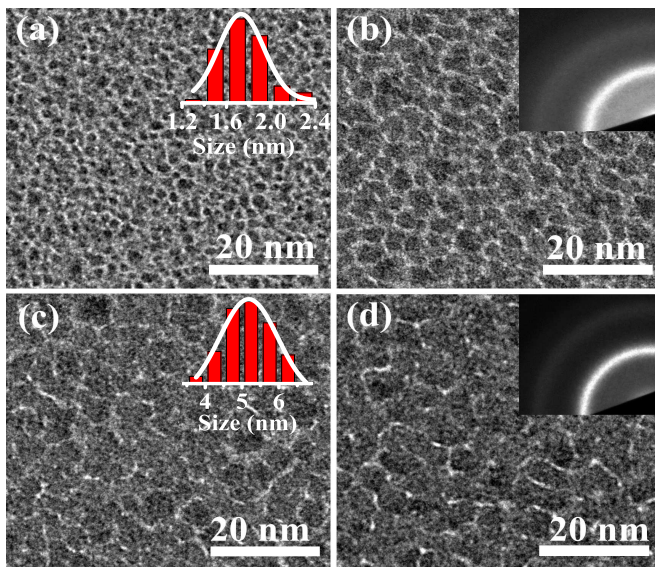}
\caption{Bright-field TEM images for Ta$_x$(SiO$_2$)$_{1-x}$ films with $x$ values of (a) 0.45, (b) 0.66, (c) 0.75, and (d) 0.8. The insets in (a) and (c) are the grain-size distribution histograms for the $x\simeq0.45$ and 0.75 films, respectively, and in (b) and (d) are the selected-area electron-diffraction patterns of the corresponding films.}\label{LiTEM}
\end{center}
\end{figure}

\section{Experimental Method}\label{SecEM}
Our Ta$_x$(SiO$_2$)$_{1-x}$ films were deposited on Al$_2$O$_3$ single crystal substrates by co-sputtering Ta and SiO$_2$ targets in Ar atmosphere. Al$_2$O$_3$ single crystal possesses relative higher thermal conductivity, which is favorable for the thermal equilibrium between the film and sample holder in liquid helium temperatures for electrical transport measurement. The base pressure of the chamber is less than $8\times 10^{-5}$\,Pa, and the deposition was carried out in an argon (99.999\%) atmosphere of 0.5\,Pa. During deposition, the substrate temperature was kept at 200\,$^\circ$C and the Ta volume fraction $x$ was regulated by the sputtering powers applied in the two targets. The films were also simultaneously deposited on the polyimide (Kapton) and copper grids
coated with ultrathin carbon films for composition and microstructure measurements. The narrow rectangle shape films (2\,mm$\times$10\,mm), defined by mechanical masks, were used for transport measurement. To obtain good contact, four Ti/Au electrodes were deposited on the films. The distance between the two voltage Ti/Au electrodes is 2\,mm.

The thickness of the films was controlled by depositing time and the exact thickness of the films was determined by the atomic force microscopy (AFM Multimode-8, Brucker). It is indicated that the thicknesses of the films are $9.0\pm 1.2$\,nm. The Ta volume fraction $x$ in each film was obtained from energy-dispersive x-ray spectroscopy analysis (EDS; EDAX, model Apollo X). The microstructure of the films was characterized by transmission electron microscopy (TEM, Tecnai G2 F20). The resistance versus temperature and magnetic field was measured using standard four probe ac technology, in which a Keithley 6221 and a SR 830 lock-in amplifier were used as the ac current source and voltmeter, respectively. During measurements, the frequency of the current is set as 13.33\,Hz and the applied bias varies from 50 to 100\,nA for different films. For the isothermal current-voltage curves measurement, the current was supplied by the Keithley 6221 and the voltage was monitored by a Keithley 2182A. The low temperature and magnetic field environments were provided by dilution refrigerator (Triton 200, Oxford).

\section{Results and Discussions}
Figure~\ref{LiTEM} shows the bright-field TEM images and selected
area electron diffraction (SAED) patterns for films with $x\simeq0.45$, 0.66, 0.75 and 0.80. The dark regions are Ta granules and the bright regions are SiO$_2$ insulating matrix. Thus these films reveal typical granular composite characteristics. The mean-size of Ta granules (d) for the $x\simeq0.45$, 0.66, 0.75 and 0.80 films is $\sim$1.8, $\sim$3.6, $\sim$5.1, and $\sim$5.6\,nm, respectively (see the insets in Figs.~\ref{LiTEM})(a) and \ref{LiTEM}(c). Only two dispersive diffraction rings present in the SAED patterns, indicating that both Ta grains and SiO$_2$ matrix are amorphous.

\begin{figure}
\begin{center}
\includegraphics[scale=0.9]{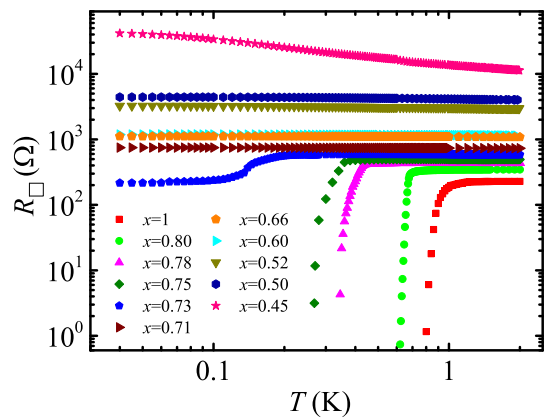}
\caption{Sheet resistance as a function of temperature from 2 down to 0.04\,K for the Ta$_x$(SiO$_2$)$_{1-x}$ films ($0.45\lesssim x \leq 1$).}\label{LiRT2-40mK}
\end{center}
\end{figure}

Figure~\ref{LiRT2-40mK} shows temperature $T$ dependence of the sheet resistance $R_\square$ for films with $0.45\lesssim x \lesssim 0.80$ from 2\,K down to 40\,mK. Clearly, the normal-state sheet resistance $R_{N}$ decreases with increasing $x$ at a certain temperature. For the $x\gtrsim0.75$ films, the low-temperature sheet resistance changes little with decreasing temperature in the normal state, and then rapidly drops to zero below a transition temperature $T_c$ with further decreasing temperature, where the superconducting transition temperature $T_c$ is defined as the temperature at which the sheet resistance drops to $0.9R_{N}$. The value of $T_c$ decreases with reducing $x$, and is listed in Table~\ref{TableI}. For the $x\simeq0.73$ film, the sheet resistance drops from $R_{N}$ to $\sim$0.41$R_{N}$ as the temperature decreases from 0.25 to 0.1\,K, and then tends to be a constant with further decreasing temperature. Thus the the anomalous metallic state is achieved in the Ta$_x$(SiO$_2$)$_{1-x}$ granular films via tuning the volume fraction of Ta, i.e., tuning the disorder of the system. For the $0.50\lesssim x\lesssim 0.71$ films, a very slow increase in  $R_\square$ is visible upon cooling, e.g., the variation of $R_\square$ is less than 10\% when the films is cooled from 2\,K down to 40\,mK. For the $x\simeq 0.45$ film, the sheet resistance increases a factor of 4 in this temperature range, indicating the film has transformed into insulator. We will focus on the $x\gtrsim 0.73$ films in the following discussions.

\begin{figure}
\begin{center}
\includegraphics[scale=1.1]{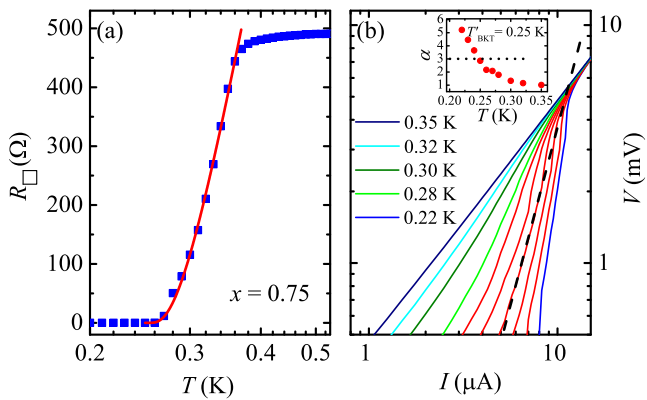}
\caption{(a) The sheet resistance $R_\square$ versus temperature $T$ for the $x \simeq 0.75$ film. The
solid curve is the least-squares fits to the Halperin-Nelson formula. (b) Voltage versus current (in double logarithmic scales) measured at constant
temperature ranging from 0.22 to 0.35\,K for zero magnetic field, the steps are 0.01\,K from 0.22 to 0.28\,K. Inset: $\alpha$ as a function of $T$. The dashed line represents the $\alpha=3$ line.}\label{FIGBKT}
\end{center}
\end{figure}

\subsection{Berezinskii-Kosterlitz-Thouless transition}

A remarkable feature for 2D superconducting films, no matter homogeneous or inhomogeneous, is the occurrence of Berezinskii-Kosterlitz-Thouless (BKT) transition~\cite{no31,no32,no33,no34}, which in turn becomes a criterion to judge whether a superconductor is 2D. The basic picture of the BKT transition is the existence of thermally excited vortices which are bound in vortex-antivortex pairs below the phase transition temperature $T_{\rm BKT}$ and dissociated above. In Fig.~\ref{FIGBKT}(a), we present $R_\square$ variation with $T$ for the $x\simeq 0.75$ film from 0.5 down to 0.2 K. It is found that the temperature dependence of the sheet resistance in the superconducting transition region can be well described by the Halperin-Nelson formula~\cite{no35} $R_\square=R_0\exp[-b(T/T_{\rm BKT}-1)^{-1/2}]$, where $R_0$ is a prefactor, $b\sim 1$ is a constant, and $T_{\rm BKT}$ is the BKT transition temperature, which can be determined via extrapolating the linear part of $[{\rm d}\ln R_\square/{\rm d}T]^{-2/3}$ versus $T$ curve to $[{\rm d}\ln R_\square/{\rm d}T]^{-2/3}=0$. For the $x\simeq 0.75$ film, the value of $T_{\rm BKT}$ is $\sim$0.25 K, and the solid curve in Fig.~\ref{FIGBKT}(a) is the least-squares fit to the Halperin-Nelson formula with $R_0=7215 $\,$\Omega$ and $b=0.92$. On the other hand, in the BKT transition region the current dependence of voltage obeys $V\sim I^{\alpha(T)}$ law in small current limit, and the value of $\alpha(T)$ is $\alpha(T)=3$ at $T_{\rm BKT}$. These relations valid for both homogeneous and inhomogeneous (granular films or proximity-coupled arrays) 2D superconductors~\cite{no31,no32,no33,no34}. Figure~\ref{FIGBKT}(b) shows the voltage versus current at some selected temperatures in double logarithmic scales for the $x\simeq 0.75$ film. Clearly, $\log V$ varies linearly with $\log I$ in small current limit, indicating the relation $V\sim I^{\alpha}$ is satisfied. The inset of Fig.~\ref{FIGBKT}(b) shows $\alpha$ as a function of $T$ obtained via linear fitting of the $\log V$-$\log I$ data. Inspection of the figure indicates that $\alpha(T)$ decreases from $\sim$5.2 to 1 as the film is heated from 0.22 to 0.35\,K, and the BKT transition temperature is $T_{\rm BKT}\simeq 0.25$\,K, which is identical to that determined by the Halperin-Nelson formula within the experimental uncertainty. Similar phenomena have also been observed in the $x>0.75$ films, and the transition temperature $T_{\rm BKT}$, together with the fitting parameters $R_0$ and $b$, is listed in Table~\ref{TableI}. The existence of BKT transition in the $x\gtrsim 0.75$ films suggests that these films are 2D with respect to superconductivity.

\begin{figure}
\begin{center}
\includegraphics[scale=1.1]{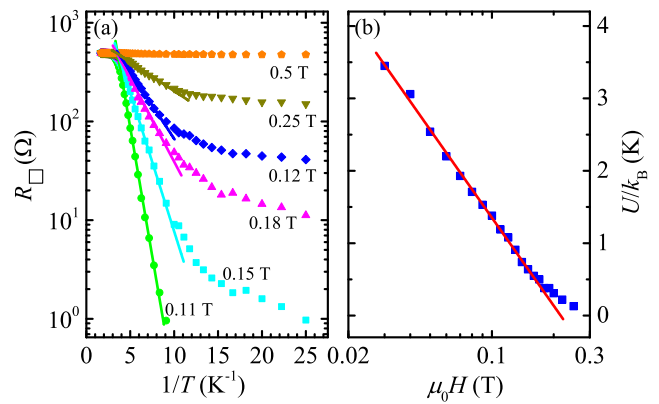}
\caption{(a) Logarithm of resistivity as a function of $T^{-1}$ at different magnetic field for the $x$ $\simeq$ 0.75 films. The symbols are the experimental data and the straight solid lines are least-squares fits to $R_\square =R_0(H)\exp(-U(H)/k_BT)$. (b) Activation energy $U(H)/k_B$, obtained from the slopes of the solid lines in Fig.~\ref{FigHeR-TR}(a), as a function magnetic field for the $x$ $\simeq$ 0.75 film. The solid line is the least-squares fit to Eq.~(\ref{Eq-TAC}). }\label{FigHeR-TR}
\end{center}
\end{figure}

\begin{table*}
\caption{\label{TableI} Relevant parameters for the Ta$_x$(SiO$_2$)$_{1-x}$ films with $x\gtrsim 0.73$. Here $x$ is volume fraction of Tb,  $T_c$ is the
superconducing transition temperature, $T_{\rm BKT}$ is the BKT transition temperature, $R_0$ and $b$ are the parameters in Halperin-Nelson formula, $U_0$ and $H_0$ are the parameters in Eq.~(\ref{Eq-TAC}), $H_{c2}(0)$ is upper critical field at 0\,K, $d$ is the the mean-size of Ta granules, $\xi(0)$ is the Ginzburg-Landau coherence length at zero-temperature, and $H_c$ is the critical field at which the isotherms of the $R_\square$ versus $\mu_0 H$ cross at a point.}
\begin{ruledtabular}
\begin{center}
\begin{tabular}{cccccccccccc}
       & $R_\square(2\,{\rm K})$    & $T_c$    & $T_{\rm BKT}$  & $d$ &$R_{0}$   & $b$   &  $U_0/k_B$      &  $H_0$   &   $H_{c2}(0)$       & $\xi(0)$ & $H_c$ \\
$x$    &  ($\Omega$)  &(K)     &  (K) &  (nm) & ($\Omega$) &       &   (K)  & (T)  &  (T) &    (nm) & (T)   \\  \hline
1.0    &   227    & 0.91   &   0.82 & $-$ & 667  &  1.04  & 7.22            &    1.06    &   1.51         &  14.5 &  1.80   \\
0.80   &   345    & 0.66   &   0.62 & 5.6 & 41534  &  1.18   &   3.67      &    0.69    &   0.96         &  18.2  &  1.22  \\
0.78   &   434    & 0.39   &  0.34 & 5.3 & 4520  &  0.78   &   1.26      &    0.39    &   0.52         &  24.7   & 0.71\\
0.75   &   493    & 0.32   &  0.25 & 5.1 & 7215  &  0.92   &   1.75      &    0.22    &   0.45         &  26.6   & 0.65\\
0.73   &   585    & $-$    &   $-$  & 4.7 & $-$   &  $-$    &   $-$       &      $-$   &   0.11          &  53.8   \\
\end{tabular}
\end{center}
\end{ruledtabular}
\end{table*}

\subsection{Anomalous metallic state}
For the $x\gtrsim 0.75$ films, the superconductivity would gradually disappear once a moderate magnetic field perpendicular to the film plane was applied. Considering the temperature dependence behaviors of the $R_\square$ under magnetic field are similar for the $x\gtrsim 0.75$ films, we only present and discuss the results obtained in the $x \simeq 0.75$ film. Figure~\ref{FigHeR-TR}(a) shows the $R_\square$ (in logarithmic scale) as a function of $1/T$ at different magnetic fields, as indicated. The low-temperature state of the film remains superconductivity (zero resistance) at low field.  When a moderate field with magnitude $H_{c0}\lesssim H \lesssim H_{c2}$ ($H_{c2}$ is upper critical magnetic field, while $H_{c0}$ is the critical magnetic field for superconductor to anomalous metal transition which will be defined later) is applied, the sheet resistance starts to drop rapidly with decreasing temperature near $T_c$, and then tends to saturate with further decreasing temperature. The saturation value increases with the enhancement of the field until the field reaches $H_{c2}$. The values of $\mu_0 H_{c2}$ (with $\mu_0$ being the permeability of free space) for the $x\gtrsim 0.75$ film are summarized in Table~\ref{TableI}.  The features of $R_\square$ suggest that an anomalous metallic ground state presents in these $x\gtrsim 0.75$ films when a moderate field is applied.

From Fig.~\ref{FigHeR-TR}(a), one can see that the $\log R_\square$ ($\ln R_\square$) decreases linearly with increasing $1/T$ in the transition region near $T_c$. Thus $R_\square(T)$ satisfies $R_\square =R_0(H)\exp(-U(H)/k_BT)$ in the transition region, where $R_0(H)$ is a prefactor, $k_B$ is the Boltzmann constant, and $U(H)$ is the activation energy under field $H$. This form of $R_\square(T)$ means that the vortex-antivortex pairs are unbound with increasing temperature and the resistance in the transition region is governed by the motion of thermal activated individual vortices. Figure~\ref{FigHeR-TR}(b) shows the activation energy $U(H)$ variation with $H$ extracted from least-squares fits to the linear part of the $\log R_\square$ versus $1/T$ data in Fig.~\ref{FigHeR-TR}(a). Clearly, the activation energy $U(H)$ variation with $H$ is consistent with the thermally assisted collective vortex-creep model in 2D system~\cite{no36},
\begin{equation}\label{Eq-TAC}
U(H)=U_0\ln(H_0/H)
\end{equation}
where $U_0$ is the vortex-antivortex binding energy and $H_0\sim H_{c2}(0)$ is the field above which all the vortex-antivortex pairs are almost broken down. The fitted values of $U_0$ and $H_0$ are listed in Table~\ref{TableI}. The value of $H_{c2}(0)$ can be estimated by linear extrapolating the low temperature $H_{c2}(T)$ data to 0\,K. Thus the in-plane Ginzburg-Landau coherence length at zero temperature $\xi(0)$ is obtained via $\mu_0 H_{c2}(0)=\Phi_0/2\pi\xi^2(0)$ with $\Phi_0$ being flux quantum. The values of $H_{c2}(0)$ and $\xi(0)$ are also summarized in Table~\ref{TableI}.  Insepectation of Table~\ref{TableI} indicates the value of $H_0$ is less than the upper critical field $H_{c2}(0)$ for each film. Since the thermally assisted collective vortex-creep model is heuristic~\cite{no36}, the deviation is acceptable. The coherence length $\xi(0)$ is greater than the thickness $t$ for each film, which confirms the 2D superconductor characteristics of the films.

\begin{figure}
\begin{center}
\includegraphics[scale=1.1]{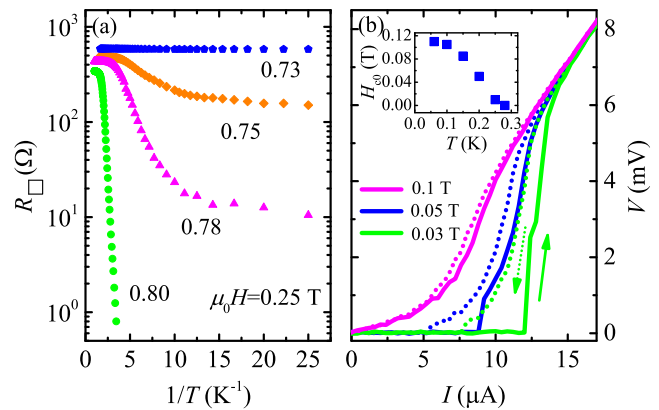}
\caption{(a) Logarithm of the sheet resistance as a function of $T^{-1}$ from 1 down to 0.04\,K at a field of 0.25\,T for the $x$$\simeq$$0.73,\, 0.75,\, 0.78$ and  0.80 films. (b) $I$-$V$ curves for the $x\simeq 0.75$ film measured at different fields. Inset: $H_{c0}$ as a function of temperature for the $x\simeq 0.75$ film.}\label{FigHeI-V}
\end{center}
\end{figure}

Figure~\ref{FigHeI-V}(a) shows the $R_\square$ as a function of $1/T$ at 0.25\,T for the $0.73\lesssim x\lesssim 0.80$ films. The $x \geq 0.8$ films remain superconducting, while the $0.73\lesssim x < 0.80$ films exhibit metallic behavior below $\sim$$0.1$\,K. The immediate effect of decreasing $x$ on the films is the enhancement of disorder. Thus the anomalous metallic state in Ta$_x$(SiO$_2$)$_{1-x}$ granular films can also be driven by disorder besides magnetic field. The reduction of $x$ decreases the mean-size of Ta granules and increases the separation between Ta granules, thus increases the tunneling resistance, which results in the enhancement of the superconducting quantum phase fluctuations at low temperatures~\cite{no37}. In addition, the $T_c$ of superconducting particles with intermediate electron-phonon coupling strength (i.e., the $T_c$ and superconducting energy gap $\Delta(0)$ of the bulk superconductor satisfies $2\Delta(0)\sim 3.5 k_B T_c$) decreases with decreasing particle size due to quantum size effect~\cite{no38,no39}. The value of $2\Delta/k_BT_c$ of bulk Ta is $\sim$3.5~\cite{no40}, slightly less than that of Nb ($\sim$3.9)~\cite{no39}. Thus the magnitude of $\Delta(0)$ (or $T_c$) of Ta particles will decrease with decreasing particle size (or $x$ in granular films), which is just what we observe in the Ta$_x$(SiO$_2$)$_{1-x}$ granular films. For the Ta$_x$(SiO$_2$)$_{1-x}$ granular films with a certain fixed $x$, the numbers of grains of Ta with a certain size versus the grain size tends to follow a normal distribution (see insets in Fig.~\ref{LiTEM}). Thus the values of $\Delta(0)$ for Ta grains also obey the normal distribution law, which could enhance the fluctuations of the superconductor order parameter. On the other hand, the external magnetic field not only reduces the superconductor energy gap but also destroy the coherence between the superconducting grains, which also increases the superconducting quantum phase fluctuation. Thus the quantum fluctuations of the superconductor order parameter plays a key role in the superconductor to anomalous metal transition.

Figure~\ref{FigHeI-V}(b) shows the voltage $V$ as a function of current $I$ at 60 mK at different field for the $x\simeq 0.75$ film. In the larger current part, the $I$-$V$ curve exhibits nonlinear behavior, and the curve measured in the increasing current process does not overlap with that in the decreasing process at a moderate field. As the field increases, the hysteresis in $I$-$V$ curve gradually disappears. The minimum field at which the hysteresis disappears is defined as the critical field $H_{c0}$ for superconductor to anomalous metal transition~\cite{no3,no4,no9}. $H_{c0}$ is temperature dependent and shown in the inset of Fig.~\ref{FigHeI-V}(b). For the $x\simeq 0.75$ film, $H_{c0}$ decreases from 0.11 to 0.04\,T as $T$ increases from 0.04 to 0.2\,K.  The hysteretic in $I$-$V$ curves was also observed in homogeneous Ta film~\cite{no3,no4}, 2D crystalline NbSe$_2$ films~\cite{no9} and granular Bi film~\cite{no41} under a magnetic field. The characteristics of $I$-$V$ curves in our films are similar to that in the  Josephson junction in the underdamped region~\cite{no43}, which is consistent with the granular structure of our films.

\begin{figure}
\begin{center}
\includegraphics[scale=1.1]{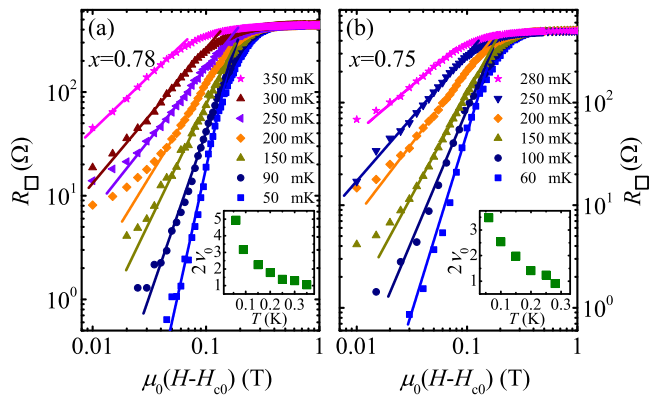}
\caption{Sheet resistance as a function of $H$-$H_{c0}$ in double logarithmic scales at different temperatures for (a) $x \simeq 0.78$, and (b) $x\simeq 0.75$ films. The solid lines are least-squares fits to Eq.~(\ref{Eq-ScalingSM}). Inset: The exponent $2\nu_0$ in Eq.~(\ref{Eq-ScalingSM}) as a function of $T$ for the corresponding films.  }\label{FigHeScaling-1}
\end{center}
\end{figure}

\subsection{The scaling relation for the resistance near the phase transitions }
Das and Doniach have proposed a Bose-metal model to explain the anomalous metallic state in 2D superconductors~\cite{no22,no23}. They investigated the quantum phase fluctuations in granular superconductors in the absence of disorder. In their picture, the anomalous metal phase is dominated by dynamical gauge field fluctuations which results from superconducting quantum fluctuations. Vortices and antivortices moving in a dynamically fluctuating gauge field tend to form a quantum liquid when gauge field fluctuations overcome the quantum zero point motion of the vortices. They argue that the ground state of the system are determined by three energy: the Josephson-coupling energy $J$, the onsite repulsion energy $V_0$, and repulsion energy among the nearest neighbors $V_1$ ($V_0$ and $V_1$ are related to the inverse of the capacitance matrix of the grains). When $V_0/J+8V_1/J$ is less than $\tilde{b}_0$ ($\tilde{b}_0\simeq16.78$ in Ref. 22), the system is superconducting; when $V_0/J+8V_1/J$ is greater than $\sim$$16\sqrt{3V_1 /J}$, the system is insulaing; and in the intermediate region, the system is in Bose-metal state. In the granular superconductor films, $J=(R_Q/2R_N)\Delta$ ($R_Q=h/4e^2$ with $h$ being the Planck's constant and $e$ the electron charge) decreases with decreasing $x$, and a crossover from superconducing to Bose-metal states would occur as $V_0$ and $V_1$ are of comparable order of magnitude. This is just what we see in the $x\simeq 0.73$ film at zero field.

When a small magnetic field is applied, the coherence of the superconducting region will be gradually suppressed and the coupling energy $J$ will be reduced. The uncondensed bosons (vortex and antivortex) start to emerge at $H>H_{c0}$, which drives the system into the Bose-metal phase.  Thus the superconductor to Bose-metal transition is associated with the unbinding of (quantum) dislocation-antidislocation (or vortex and antivortex) pairs and finite resistance is induced by the free dislocations (vortices). Near the superconductor to Bose-metal transition and on the metallic side, the resistance scales with~\cite{no23}
\begin{equation}\label{Eq-ScalingSM}
R_\square\sim (H-H_{c0})^{2\nu_0},
\end{equation}
where $H_{c0}$ is the critical field for superconductor to Bose-metal transition and $\nu_0$ is the scaling parameter. As the field increases further, quantum zero point motion of the vortices increases, and it gradually overtakes the dynamical gauge field fluctuations. When the field is greater than a critical value $H_c$, quantum zero point motion of the vortices dominates the vortex dynamics. The vortices form a superfluid phase and the film is insulating. Considering the insulator to Bose-metal transition is a phase transition from
a vortex superfluid to a gapless nonsuperfluid phase, Das and Doniach proposed a two parameter scaling relation for resistance across the transition~\cite{no23}
\begin{equation}\label{Eq-ScalingMI}
R_\square\left[\frac{T^{1/\nu z}}{\delta}\right]^{\nu(z+2)}=f(\delta/T^{1/\nu z})
\end{equation}
where $\delta=\mu_0(H-H_c)$ with $H_c$ being the critical field for metal-insulator transition, $\nu$ and $z$ are scaling exponents. Similar to the scaling relation $R_\square\sim F(\delta/T^{1/\nu z})$ near the superconductor to insulator transition in disordered 2D superconductors~\cite{no43}, Eq.~(\ref{Eq-ScalingMI}) was derived basing a second order quantum phase transition. Near the transition there is a diverging correlation length $\xi\sim |\delta|^{-\nu}$ and a vanishing frequency characteristic frequency $\Omega\sim\xi^{-z}$. Das and Doniach compare the experimental data of MoGe films with Eq.~(\ref{Eq-ScalingMI}) and find the data collapse onto two different branches with $\nu=4/3$ and $z=1$ for $H>H_c$ and $H<H_c$.

\begin{figure}
\begin{center}
\includegraphics[scale=0.8]{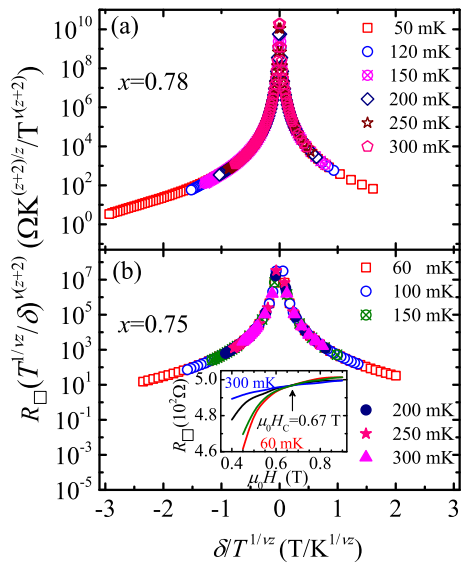}
\caption{(a) $R_\square({T^{1/\nu z}/\delta})^{\nu(z+2)}$ as a function of $\delta/T^{1/\nu z}$ at different temperatures and fields near the anomalous
metal to insulator transition for (a) $x$ $\simeq$ 0.78, and (b) $x$ $\simeq$ $0.75$ films. Here $\nu z$ is taken as 4/3 in both (a) and (b). The magnetic field data  are taken from 0.4 to 0.9\,T for both films at a certain fixed temperature. The inset in (b) is $R_\square$ versus $\mu_0 H$ measured at different temperatures (300, 200, 100, and 60\,mK from up to down on the left side of the cross point) for the corresponding film. }\label{FigHeScaling-2}
\end{center}
\end{figure}

Figure~\ref{FigHeScaling-1} (a) and \ref{FigHeScaling-1}(b) show $R_\square$ as a function of $\mu_0(H-H_{c0})$ in double-logarithmic scales for the $x \simeq 0.78$ and 0.75 films, respectively. Clearly, $\log R_\square$ varies linearly with $\log\mu_0(H-H_{c0})$ in the vicinity of $H_{c0}$, indicating the scaling relation in Eq.~(\ref{Eq-ScalingSM}) is present in the films. The linear parts of the curves are least-squares fitted to Eq.~(\ref{Eq-ScalingSM}), and the results are shown in Fig.~\ref{FigHeScaling-1} by solid lines. The insets in Fig.~\ref{FigHeScaling-1} present temperature dependence of the exponent $2\nu_0$. For the $x \simeq 0.78$ (0.75) film, the value of $2\nu_0$ gradually decreases from $\sim$4.94 (3.48) to 1.05 (0.91) as the temperature increases from $0.04$\,K to higher temperature. $2\nu_0\simeq 1$ in Eq.~(\ref{Eq-ScalingSM}) means a rough linear field dependence of resistance, indicating unhindered flux flow gradually dominates the transport process in the films at higher temperatures.

When the field is much larger than $H_{c0}(T)$, the scaling law in Eq.~(\ref{Eq-ScalingSM}) is not valid. According to the theoretical prediction of Bose-metal model, Eq.~(\ref{Eq-ScalingMI}) should be satisfied in the vicinity of Bose-metal to insulator transition. Figure~\ref{FigHeScaling-2} shows $R_\square({T^{1/\nu z}/\delta})^{\nu(z+2)}$ as a function of $\delta/T^{1/\nu z}$ near the transition with $\nu z=4/3$. The inset of Fig.~\ref{FigHeScaling-2}(b) presents $R_\square$ as a function of $\mu_0 H$ measured at different temperatures for the $x\simeq 0.75$ film. The isotherms of $R_\square$ versus curves cross at a critical magnetic field $H_c$, at which the critical sheet resistance $R_{\square}^{c}$ is $\sim$495\,$\Omega$. The value of $R_{\square}^{c}$ decreases from $\sim$495 to $\sim$231\,$\Omega$ as $x$ increases from $\sim$0.75 to 1, and the value of $H_c$ for each film is listed in Table~\ref{TableI}. Figure~\ref{FigHeScaling-2} clearly indicates that the $R_\square (H,T)$ data at  $H<H_c$ and $H>H_c$ collapse onto two different branches. The value of $z$ is set as $1$ due to the long range Coulomb interaction between the Bosons in 2D system. We also compare our $R_\square (H,T)$ data with the scaling relation $R_\square\sim F(\delta/T^{1/\nu z})$ (see Supplemental Materials Fig.S1). The value of $\nu$, determined by the method in Ref~\cite{no44}, is $1.30$ for $x\simeq 0.78$ film and $1.22$ for $x\simeq 0.75$ film. It is found that the formula works at high temperature regime, but fails to describe the low-temperature $R_\square (H,T)$ data near $H_c$. Thus the validation of both Eqs.~(\ref{Eq-ScalingSM}) and (\ref{Eq-ScalingMI}) in Ta$_x$(SiO$_2$)$_{1-x}$ ($x\gtrsim 0.75$) films strongly suggests the intermediate metallic state in this 2D granular composite is the Bose-metal phase.

\begin{figure}
\begin{center}
\includegraphics[scale=1.1]{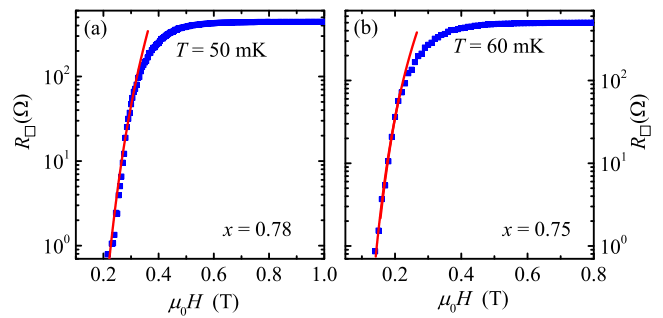}
\caption{$R_\square$ as a function of magnetic field  for (a)  $x$$\simeq$$0.78$ film at 60\,mK, and (b) $x$$\simeq$$0.75$ film at 50\,mK. The solid curves are the least-squares fits to Eq. (4).}\label{FigHe-VQT}
\end{center}
\end{figure}

\subsection{Comparison with other models}
Shimshoni et al ascribe the dissipation in 2D disordered superconductor to quantum tunneling of vortices through thin superconducting constrictions~\cite{no25}. In this scenario, the system is treated as a percolation network coupling with a fermionic bath and the resistance in the anomalous metallic state is caused by the vortices tunneling across the narrow superconducting channel. Then the low-temperature resistance at a field $H$ can be written as~\cite{no7,no25}
\begin{equation}\label{Eq-QTV}
R_\square\sim\frac{h}{4e^2}\frac{\kappa}{1-\kappa}
\end{equation}
with
\begin{displaymath}
\kappa\sim\exp\left[C\frac{h}{4e^2}\frac{1}{R_N}\left(\frac{H-H_{c2}}{H}\right) \right],
\end{displaymath}
where $C$ is a dimensionless constant of order unity, and $R_N$ is the normal state resistance.
It has been found the experimental data on amorphous MoGe films~\cite{no1,no2} and ZrNCl~\cite{no7} electric-double-layer transistor can be well described by Eq.~(\ref{Eq-QTV}).
Figure~\ref{FigHe-VQT} presents the low temperature isothermal $R_\square$-$\mu_0 H$ curves for the $x\simeq0.78$ and 0.75 films, as indicated. The solid curve are the best  fit (using least-squares method) to Eq.~(\ref{Eq-QTV}), in which the adjust parameter $C$ is 0.467 for the $x\simeq0.78$ film and 0.326 for the  $x\simeq0.75$ film.  Clearly,  Eq.~(\ref{Eq-QTV}) can only describe the $R_\square$-$\mu_0 H$ data at low field regime, e.g., $\mu_0 H\lesssim 0.30$\,T for the $x\simeq 0.78$ film and $\mu_0 H\lesssim 0.20$\,T for the $x\simeq 0.75$ film. The experimental $R_\square$-$\mu_0 H$ data at relative larger field region deviates to the prediction of Eq.~(\ref{Eq-QTV}). This suggests that the $R_\square(T)$ data in Ta$_x$(SiO$_2$)$_{1-x}$ films cannot be fully explained by the quantum tunneling of vortex model.

To explain the field-induced intermediate metallic phase in amorphous superconducting film, Galitski \emph{et al} introduce a two-fluid formulation consisting of fermionized field-induced vortices and electrically neutralized Bogoliubov quasiparticles (spinons) interacting via a long-ranged statistical interaction~\cite{no28}. The ``vortex metal" phase can be obtained in their theory, and the theory also predicates a large peak in low-temperature magnetoresistance. Form Fig.~\ref{FigHe-VQT}, one can see that the resistance increases monotonically with increasing magnetic field up to $H_{c2}(0)$ and no magnetoresistance peak appear over the whole magnetic field range. Thus the intermediate metallic phase in Ta$_x$(SiO$_2$)$_{1-x}$ granular films cannot be explained by the ``vortex metal" theory either.

\section{Conclusion}
The low temperature transport properties of a series of 2D Ta$_x$(SiO$_2$)$_{1-x}$ granular films are systematically investigated. At zero field, the low temperature states of the films undergo transitions from superconductor to anomalous metal and then to insulator with decreasing $x$. BKT transition is observed in those films with superconducting ground state ($x\gtrsim 0.75$). When a small perpendicular magnetic field is applied, transition from superconducting to anomalous metallic states is induced in those $x\gtrsim 0.75$ films. It is found that the experimental $R_\square (H,T)$ data near the transitions from superconducting to anomalous metallic states and anomalous metallic to insulating states are consistent with the theoretical scaling relations of the Bose-metal model, and cannot be described by the predictions of other related models. Our results suggest that the intermediate anomalous metallic state in the 2D granular films originates from strong dynamical gauge field fluctuations caused by the disorder and magnetic field.

\begin{acknowledgments}
The authors are grateful to Professor J. J. Lin (National Chiao Tung University) for valuable discussion. This work is supported by the National Natural Science Foundation of China through Grant No. 11774253 (Z.Q.L.) and Grant No.11774367 (G.L.Y.).
\end{acknowledgments}

\end{document}